\documentclass[a4paper]{jpconf}
\usepackage{graphicx}
\usepackage[pass]{geometry}
\usepackage{graphicx}
\usepackage{amsmath}
\usepackage[utf8]{inputenc}
\usepackage{amsmath}
\usepackage{array}
\usepackage{amsfonts}
\usepackage{epsf}
\usepackage{epsfig}
\usepackage{amssymb}
\usepackage{bbm}
\usepackage{delarray}
\begin{document}
\title{Relativistic Approximate Solutions for a Two-Term Potential: Riemann-Type Equation}

\author{Altug Arda}

\address{Department of
Physics Education, Hacettepe University, 06800, Ankara, Turkey}

\ead{arda@hacettepe.edu.tr}

\begin{abstract}
Approximate analytical solutions of a two-term potential are studied for the relativistic wave equations, namely, for the Klein-Gordon and Dirac equations. The results are obtained by solving of a Riemann-type equation whose solution can be written in terms of hypergeometric function $\,_{2}F_{1}(a,b;c;z)$. The energy eigenvalue equations and the corresponding normalized wave functions are given both for two wave equations. The results for some special cases including the Manning-Rosen potential, the Hulth\'{e}n potential and the Coulomb potential are also discussed by setting the parameters as required.
\end{abstract}

\section{Introduction}
The investigation of the non-relativistic and relativistic bound/scattering state solutions of exponential-type potentials has become an important area within quantum mechanics. For last decades, many authors have made much efforts to obtain the analytical solutions of the exponential-type potentials, especially about the problems based on the Morse potential, Hulth\'{e}n potential, and the Woods-Saxon potential, with the help of various methods [1-32]. Within this context, we deal with the following two-term exponential potential
\begin{eqnarray}
V(r)=-V_{0}\,\frac{e^{-\beta r}}{1-qe^{-\beta
r}}+V_{1}\,\frac{e^{-2\beta r}}{(1-qe^{-\beta r})^2}\,,
\end{eqnarray}
which firstly proposed in [33]. Jia \textit{et. al.} have studied the bounded solutions of the Schrödinger equation for this potential by using SUSY-approach [34], and Benarami \textit{et. al.} have investigated the same problem in terms of Green's function [35]. Arda and Sever have also presented the approximate, analytical solutions of the above potential within functional analysis method  for $q \geq 1$ and $q=0$ [36].

In the present work, we tend to give the approximate, analytical solutions of the Klein-Gordon and Dirac equations which will be written in the form of a Riemann-type equation. We write the eigenvalue equations and the corresponding "wave functions" both for the above equations with the two-term exponential potential. We present also briefly the results for the Manning-Rosen potential, the Hulth\'{e}n potential and Coulomb potential by setting the potential parameters as required. The readers can find an application of the Riemann equation [37] within the quantum mechanics in [27] where the bounded and scattering state solutions of the Klein-Gordon and Dirac equations have been studied for an exponential potential.

\section{Analytical Solutions}

\subsection{Klein-Gordon Equation}

Time-independent part of the Klein-Gordon equation in the presence of scalar, $S(r)$, and vector, $V(r)$, potentials is written [37]
\begin{eqnarray}
\left[\hbar^2c^2\vec{\nabla}^{2}+(E-V(r))^{2}-(m_{0}c^2+S(r))^2\right]\Psi(\vec{r})=0\,,
\end{eqnarray}
for a spinless particle with mass $m_{0}$ and energy $E$, where $c$ is speed of light. Since the potentials in (2) are spherical symmetric, we write the radial part of (2) by taking $\Psi(\vec{r})=R(r)Y(\theta,\phi)$ 
\begin{eqnarray}
\frac{1}{r^2}\frac{d}{dr}\left(r^2\frac{dR(r)}{dr}\right)+\frac{1}{\hbar^2c^2}\left[(E-V(r))^{2}-(m_{0}c^2+S(r))^2-\frac{C}{r^2}\right]R(r)=0\,,
\end{eqnarray}
where we have to write the parameter $C$ in terms of the angular momentum quantum number $\ell$ as $C=\ell(\ell+1)$ because of the mathematical restrictions [37].

For the case where the scalar potential is equal to the vector potential, we get from (3)
\begin{eqnarray}
\left[\frac{d^2}{dr^2}-\frac{\ell(\ell+1)}{r^2}+Q^2(E^2-m_{0}^{2}c^{4})-2Q^2(E+m_{0}c^2)V(r)\right]u(r)=0\,.
\end{eqnarray}
which is obtained by writing $R(r)=\frac{u(r)}{r}$ with $Q^2=1/\hbar^2c^2$.

By inserting two-term potential (1) into (4), using the approximation instead of the centrifugal term $\frac{1}{r^2} \sim \beta^2\,\frac{e^{-\beta r}}{(1-qe^{-\beta r})^2}$,
and defining a new variable as $z/q=e^{-\beta r}$ in (4) ($1 \leq z \leq 0$ for $0 \leq r < \infty$ with $q \rightarrow 1$), we obtain
\begin{eqnarray}
\frac{d^2u(z)}{dz^2}+\frac{1}{z}\frac{du(z)}{dz}+\left[-\frac{A_{1}^{2}}{z}+A_{2}^{2}-\frac{A_{3}^{2}}{q}\frac{1}{1-z}\right]\frac{u(z)}{z(1-z)}=0\,,
\end{eqnarray}
where
\begin{eqnarray}
A_{1}^{2}=\frac{Q^2}{\beta^2}(m_{0}^{2}c^{4}-E^2)\,;A_{2}^{2}=A_{1}^{2}+\frac{2Q^2}{q\beta^2}(E+m_{0}c^2)\left(V_{0}+\frac{V_{1}}{q}\right)\,;
A_{3}^{2}=\ell(\ell+1)+\frac{2Q^2V_{1}}{q\beta^2}(E+m_{0}c^2)\,.
\end{eqnarray}

Eq. (5) is a special form of the Riemann equation having regular singular points at $\rho, \sigma$ and $\tau$ [38, 39]
\begin{eqnarray}
\frac{d^2y(x)}{dx^2}&+&\left(\frac{1-a-a'}{x-\rho}+\frac{1-b-b'}{x-\sigma}+\frac{1-c-c'}{x-\tau}\right)\frac{dy(x)}{dx}\nonumber\\&+&
\left[\frac{aa'(\rho-\sigma)(\rho-\tau)}{x-\rho}+\frac{bb'(\sigma-\tau)(\sigma-\rho)}{x-\sigma}+\frac{cc'(\tau-\rho)(\tau-\sigma)}{x-\tau}\right]\nonumber\\&\times&
\frac{y(x)}{(x-\rho)(x-\sigma)(x-\tau)}=0\,,
\end{eqnarray}
which can be reduced to the following form [27]
\begin{eqnarray}
\frac{d^2y(x)}{dx^2}+\left(\frac{1-a-a'}{x}-\frac{1-c-c'}{1-x}\right)\frac{dy(x)}{dx}+\left(\frac{aa'}{x}-bb'+\frac{cc'}{1-x}\right)\frac{y(x)}{x(1-x)}=0\,.
\end{eqnarray}
where the parameters satisfy Fuch's relation as $a+a'+b+b'+c+c'-1=0$ [39]. Since the hypergeometric differential equation with solutions $\,_{2}F_{1}(p',q';r';x)$ is a special case of Riemann's equation [39], the solution of (5) can be written in terms of hypergeometric function [27]
\begin{eqnarray}
y(x)=x^{a}(1-x)^{c}\,_{2}F_{1}(p',q';r';x)\,,
\end{eqnarray}
where
\begin{eqnarray}
p'=a+b+c\,,\,\,\,\,\,q'=a+b'+c\,,\,\,\,\,\,r'=1+2a\,.
\end{eqnarray}

By comparing (5) with (8), we can set the parameters as
\begin{eqnarray}
a=+A_1\,,\,\,\,\,\,b=-A_{2}\,,\,\,\,\,\,c=1+D\,,
\end{eqnarray}
with $D=-1/2+\sqrt{1/4+A_{3}^{2}/q\,}$, and we obtain the solutions of (5) 
\begin{eqnarray}
u(z) \sim z^{A-1}(1-z)^{1+D}\,_{2}F_{1}(A_1-A_2+1+D,A_1+A_2+1+D;1+2A_1;z)\,,
\end{eqnarray}

In order to have a physical solution, we must write $A_1-A_2+1+D=-n (n=0, 1, 2, \ldots)$. This expression is the quantization condition of our system and gives the wave functions
of the Klein-Gordon equation for the two-term potential 
\begin{eqnarray}
u(z)=Nz^{A_1-1}(1-z)^{1+D}\,_{2}F_{1}(-n,-n+2A_2;1+2A_1;z)\,.
\end{eqnarray}
with the normalization constant $N$. By using (6) in quantization condition, we obtain the corresponding energy eigenvalue equation for the two-term potential 
\begin{eqnarray}
E^2-m_{0}c^2=(E+m_{0}c^2)\left(V_{0}+\frac{V_{1}}{q}\right)-
\left[\frac{\frac{Q}{q\beta}(E+m_{0}c^2)\left(V_{0}+\frac{V_{1}}{q}\right)}{n+1+D}\right]^2+\left(\frac{n+1+D}{2Q/\beta}\right)^2\,.
\end{eqnarray}

The normalization condition $\int_{0}^{\infty}|u(r)|^2dr=1$ turns into $\frac{1}{\beta}\int_{0}^{1}\frac{1}{z}|u(z)|^2dz=1$ which gives
\begin{eqnarray}
\frac{1}{\beta}N^2\int_{0}^{1}z^{2A_1-1}(1-z)^{2+2D}[\,_{2}F_{1}(-n,-n+2A_2;1+2A_1;z)]^2dz=1\,,\nonumber
\end{eqnarray}
With the help of the following expression
\begin{eqnarray}
\int_{0}^{1}z^{P'}(1-z)^{Q'}[\,_{2}F_{1}(-n,n+P'+Q'-1;P'+2;z)]^2dz=\nonumber\\
\frac{n!(n+Q'/2)\Gamma(n+Q')\Gamma(P'+1)\Gamma(P'+2)}{(n+2Q'+P'+1)\Gamma(n+P'+2)\Gamma(n+P'+Q'+1)}\,,
\end{eqnarray}
with $P'=2p''-1$, $Q'=2q''+2$ and $p''>0, q''>-3/2$, we write the normalization constant
\begin{eqnarray}
N=\left[\frac{\beta(n+A_1+D+1)\Gamma(n+2A_1+1)\Gamma(2A_2)}{n!(n+1+D)\Gamma(n+2+2D)\Gamma(2A_1)\Gamma(1+2A_1)}\right]^{1/2}\,.
\end{eqnarray}
for $q\rightarrow 1$.

We discuss briefly now the results for the special cases of the two-term potential (1) by starting with the Manning-Rosen potential. If we write the parameters in (1) as
\begin{eqnarray}
V_0=\frac{A}{2b^2}\,,\,\,V_1=\frac{\alpha(\alpha-1)}{2b^2}\,,\,\,\beta=\frac{1}{b}\,,\,\,q=1\,,
\end{eqnarray}
we obtain the above potential, and (14) gives the energy eigenvalue equation for the Klein-Gordon equation with Manninng-Rosen potential ($\hbar=c=1$)
\begin{eqnarray}
E^2-m_{0}^2=\frac{1}{2b^2}(E+m_{0})[A+\alpha(\alpha-1)]-\left[\frac{\frac{1}{2b}(E+m_{0})[A+\alpha(\alpha-1)]}{\Gamma}\right]^{2}+\frac{b^2\Gamma^{2}}{4}\,.
\end{eqnarray}
with $\Gamma=n+1/2+\sqrt{1/4+\ell(\ell+1)+(E+m)\alpha(\alpha-1)\,}$.

In order to obtain the energy eigenvalue equation for the Hulth\'{e}n potential, we set the parameters in (1) as
\begin{eqnarray}
V_1=0\,,\,\,q=1\,,
\end{eqnarray}
and (14) gives the following expression ($\hbar=c=1$)
\begin{eqnarray}
E^2-m_{0}^2=(E+m_{0})V_{0}-\left[\frac{\frac{1}{\beta}(E+m_{0})V_{0}}{n+\ell+1}\right]^2+\frac{\beta^2}{4}(n+\ell+1)^2\,.
\end{eqnarray}
The Coulomb potential could be obtained from the Hulth\'{e}n potential by taking $\beta r \ll 1$
\begin{eqnarray}
V(r)=-\frac{Ze^2}{r}\,,\nonumber
\end{eqnarray}
where we set $V_0=Ze^2\beta$. We write the energy eigenvalues of the Klein-Gordon equation for the Coulomb potential from (14) ($\hbar=c=1$)
\begin{eqnarray}
\frac{E-m_{0}}{E+m_{0}}=-\frac{Z^2e^4}{(n+\ell+1)^2}\,.
\end{eqnarray}

\subsection{Dirac Equation}
For the case of time-independent vector and scalar potentials, the Dirac equation is given [37]
\begin{eqnarray}
\{c\hat{\alpha}.\hat{p}+\hat{\beta}[m_{0}+S(r)]\}\Psi(\vec{r})=[E-V(r)]\Psi(\vec{r})\,,
\end{eqnarray}
where $m_0$ is the mass of a spin-$\frac{1}{2}$ particle, $E$ is the relativistic energy, $\hat{p}$ is the linear
momentum operator and $\hat{\alpha}$ and $\hat{\beta}$ are $4 \times 4$ matrices given by
\begin{eqnarray}
\hat{\alpha}=\begin{pmatrix} 0 & \hat{\sigma}\\ \hat{\sigma} & 0
\end{pmatrix}\,,\,\,\,\,\,\,\,\,\hat{\beta}=\begin{pmatrix} \mathbbm{1} & 0\\ 0 &
-\mathbbm{1}
\end{pmatrix}\,,
\end{eqnarray}

The Dirac spinor $\Psi(\vec{r})$ on the basis of complete set
$[\hat{H}, \hat{\kappa}, \hat{\textbf{J}}^2, \hat{\textbf{J}}_z]$ is defined as [37]
\begin{subequations}
\begin{align}
\Psi(\vec{r})&=\frac{1}{r}\begin{pmatrix}f_{n\kappa}(r)\phi_{jm}\\ig_{n\kappa}(r)\chi_{jm}\end{pmatrix} \text{for}\,\,\, \kappa=j+\frac{1}{2}\,,\\
\Psi(\vec{r})&=\frac{1}{r}\begin{pmatrix}f_{n\kappa}(r)\chi_{jm}\\ig_{n\kappa}(r)\phi_{jm}\end{pmatrix}
\text{for}\,\,\, \kappa=-(j+\frac{1}{2})\,,
\end{align}
\end{subequations}
with
\begin{eqnarray}
\phi_{jm}=\begin{pmatrix}\sqrt{\frac{j+m}{2j}\,}Y_{j-1/2,m-1/2}\\\sqrt{\frac{j-m}{2j}\,}Y_{j-1/2,m+1/2}\end{pmatrix}\,,
\,\,\,\,\,\chi_{jm}=\begin{pmatrix}\sqrt{-\frac{j-m+1}{2j+2}\,}Y_{j+1/2,m-1/2}\\\sqrt{\frac{j+m+1}{2j+2}\,}Y_{j+1/2,m+1/2}\end{pmatrix}\,,
\end{eqnarray}
where $Y_{j,m}$ are the normalized spherical harmonics [37]. In the complete set $[\hat{H}, \hat{\kappa}, \hat{\textbf{J}}^2, \hat{\textbf{J}}_z]$ of the conservative quantities, the operator $\hat{H}$ denotes the Hamiltonian of the system under consideration, $\hat{\kappa}$ describes the spin-orbit coupling operator, $\hat{\textbf{J}}$ and $\hat{\textbf{J}}_z$ are the total angular momentum and it's $z$-component operators, respectively [37]. We notice that we use the functions given in Eqs. (24a), (24b) and Eq. (25) without the subscripts in the rest of
computation.

By using Eqs. (24a) and (24b) and with the help of Eq. (22) we have two couple differential equations of first order 
\begin{subequations}
\begin{align}
\hbar c\left(\frac{d}{dr}-\frac{\kappa}{r}\right)f(r)&=[m_{0}c^2+E+S(r)-V(r)]g(r)\,,\\
\hbar
c\left(\frac{d}{dr}+\frac{\kappa}{r}\right)g(r)&=[m_{0}c^2-E+S(r)+V(r)]f(r)\,,
\end{align}
\end{subequations}
which gives us the followings for $S(r)=V(r)$
\begin{subequations}
\begin{align}
f(r)&=\frac{\hbar c}{m_{0}c^2-E+2V(r)}\left(\frac{d}{dr}+\frac{\kappa}{r}\right)g(r)\,,\\
g(r)&=\frac{\hbar
c}{m_{0}c^2+E}\left(\frac{d}{dr}-\frac{\kappa}{r}\right)f(r)\,,
\end{align}
\end{subequations}
Inserting $g(r)$ in (27b) into (27a) gives us the equation 
\begin{eqnarray}
\left[\frac{d^2}{dr^2}-\frac{\kappa(\kappa-1)}{r^2}+Q^2(E^2-m_{0}^2c^4)-2Q^2(E+m_{0}c^2)V(r)\right]f(r)=0\,.
\end{eqnarray}
which is similar to (4) with a change $\ell(\ell+1) \rightarrow \kappa(\kappa-1)$. By following the same steps in above subsection, we write the Dirac equation for the two-term potential as
\begin{eqnarray}
\frac{d^2f(z)}{dz^2}+\frac{1}{z}\frac{df(z)}{dz}+\left[-\frac{A_{1}^{2}}{z}+A_{2}^{2}-\frac{(A_{3}^{\mp})^{2}}{q}\frac{1}{1-z}\right]\frac{f(z)}{z(1-z)}=0\,,
\end{eqnarray}
with
\begin{eqnarray}
(A_{3}^{-})^{2}=\kappa(\kappa-1)+\frac{2Q^2V_{1}}{q\beta^2}(E+m_{0}c^2)\,.
\end{eqnarray}
and $(A_{3}^{+})^{2}=(A_{3}^{-})^{2}(\kappa \rightarrow \kappa+1)$. The upper indices $\mp$ of the constant $A_{3}^{2}$ is related with the upper component $f(r)$, and $g(r)$, respectively.

It is easy to obtain the energy eigenvalue equation for the Dirac equation with the two-term potential from (14), and we write
\begin{eqnarray}
E^2-E\left(V_{0}+\frac{V_{1}}{q}\right)&-&m_{0}c^2\left(V_{0}+\frac{V_{1}}{q}+1\right)+
\left[\frac{\frac{Q}{q\beta}(E+m_{0}c^2)\left(V_{0}+\frac{V_{1}}{q}\right)}{n+\frac{1}{2}+\sqrt{\frac{1}{4}+\frac{(A_{3}^{\mp})^{2}}{q}\,}}\right]^2\nonumber
\\&-&\left(\frac{n+\frac{1}{2}+\sqrt{\frac{1}{4}+\frac{(A_{3}^{\mp})^{2}}{q}\,}}{2Q/\beta}\right)^2=0\,.
\end{eqnarray}
The corresponding upper component of Dirac spinor from (13) is given as
\begin{eqnarray}
f(z)=Nz^{A_1-1}(1-z)^{1+D^{\mp}}\,_{2}F_{1}(-n,-n+2A_2;1+2A_1;z)\,,
\end{eqnarray}
with the normalization constant
\begin{eqnarray}
N=\left[\frac{\beta(n+A_1+D^{\mp}+1)\Gamma(n+2A_1+1)\Gamma(2A_2)}{n!(n+1+D^{\mp})\Gamma(n+2+2D^{\mp})\Gamma(2A_1)\Gamma(1+2A_1)}\right]^{1/2}\,.
\end{eqnarray}
with $D^{\mp}=-\frac{1}{2}+\sqrt{\frac{1}{4}+\frac{(A_{3}^{\mp})^{2}}{q}\,}$.

\section{Conclusion}
We have studied the approximate, analytical bound state solutions of the Klein-Gordon and Dirac equations for the two-term potential. We have obtained the energy eigenvalue equations, and the corresponding normalized "wave functions" by converting the wave equations to a Riemann-type equation. We have presented the results with the help of an approximation instead of the centrifugal term, and discussed the cases including the Manning-Rosen potential, the Hulth\'{e}n potential and the Coulomb potential.

\section{Acknowledgments}
This research was partially supported by Hacettepe University Scientific Research Coordination Unit, Project Code: FBB-$2016$-$9394$.

\smallskip

\end{document}